\documentclass[10pt,journal,doublecolumn]{IEEEtran}

\usepackage{threeparttable,placeins,makecell,stfloats,cite}
\usepackage{threeparttable,arydshln}
\usepackage{amsmath,amssymb,bm}
\usepackage{mathtools}
\usepackage[dvips]{graphicx}
\usepackage[usenames]{color}
\usepackage{amsfonts}
\usepackage{latexsym}
\usepackage{multirow}
\usepackage{caption}
\usepackage[labelformat=simple]{subcaption}
\usepackage{rotating}
\usepackage{float}
\usepackage{url}
\usepackage{cite}
\usepackage[noend]{algpseudocode}
\usepackage{algorithmicx,algorithm}
\usepackage[noend]{algpseudocode}
\usepackage{diagbox}
\usepackage{cuted}

\newtheorem{theorem}{\textit{Theorem}}
\newtheorem{lemma}{\textit{Lemma}}
\newtheorem{corollary}{\textit{Corollary}}

\newtheorem{observation}{\textit{Observation}}
\newtheorem{remark}{\textit{Remark}}

\usepackage{graphicx}                 
\usepackage{overpic}  
\IEEEoverridecommandlockouts

\begin{document}

\title{Generalized Arlery-Tan-Rabaste-Levenshtein\\ Lower Bounds on Ambiguity Function\\ and Their Asymptotic Achievability}

\author{Lingsheng~Meng, \textit{Graduate Student Member, IEEE}, Yong Liang~Guan, \textit{Senior Member, IEEE}, Yao~Ge, \textit{Member, IEEE}, Zilong~Liu, \textit{Senior Member, IEEE}, and Pingzhi Fan, \textit{Life Fellow, IEEE}
\thanks{Lingsheng Meng and Yong Liang Guan are with the School of Electrical and Electronics Engineering, Nanyang Technological University, Singapore 639798, and also with the Continental-NTU Corporate Lab, Nanyang Technological University, Singapore 639798 (e-mail: meng0071@e.ntu.edu.sg; eylguan@ntu.edu.sg). }
\thanks{Yao Ge is with the Continental-NTU Corporate Lab, Nanyang Technological University, Singapore 639798 (e-mail: yao.ge@ntu.edu.sg).}
\thanks{Zilong Liu is with the School of Computer Science and Electronics Engineering, University of Essex, Colchester CO4 3SQ, UK (e-mail: zilong.liu@essex.ac.uk).}
\thanks{Pingzhi Fan is with the Key Laboratory of Information Coding and Transmission of Sichuan Province, CSNMT International Cooperation Research Centre (MoST), Southwest Jiaotong University, Chengdu 610031, China (e-mail: p.fan@ieee.org).}
}

\maketitle

\begin{abstract} 
This paper presents generalized Arlery-Tan-Rabaste-Levenshtein lower bounds on the maximum aperiodic ambiguity function (AF) magnitude of unimodular sequences under certain delay-Doppler low ambiguity zones (LAZ). Our core idea is to explore the upper and lower bounds on the Frobenius norm of the weighted auto- and cross-AF matrices by introducing two weight vectors associated with the delay and Doppler shifts, respectively. As a second major contribution, we demonstrate that our derived lower bounds are asymptotically achievable with selected Chu sequence sets by analyzing their maximum auto- and cross- AF magnitudes within certain LAZ.
\end{abstract}


\section{Introduction}
\IEEEPARstart{A}{mbiguity} function (AF) is an important concept in communication, radar, and sonar systems \cite{John2007,Xiao2022,Blunt2016}. Unlike the traditional correlation functions, an AF characterizes the receiver's response to both Delay and Doppler shifts, serving as a pivotal tool for mobile wireless system design. By sending a preamble/sensing sequence over a wireless channel, for example, the AF of such a sequence can be used to measure the ranges and velocities of different targets in complex environments \cite{Skolnik2008}. 

The maximum magnitude of the AF sidelobes critically influences the target detection and delay-Doppler estimation capabilities of sequences, as the AF peaks of weaker targets may be buried by the sidelobes of strong targets. To achieve reliable detection and estimation, the ideal sequence sets are expected to have zero periodic/aperiodic auto-ambiguity function (AAF) and cross-ambiguity function (CAF) values, except for the AAF peak at the origin of the delay-Doppler plane. However, such sequence sets are impossible due to the limited volume of AFs \cite{Skolnik2008}. In practical applications, a feasible approach is to optimize the local AF sidelobes of sequences over a low ambiguity zone (LAZ) \cite{Ye2022} in the delay-Doppler plane, since the maximum delay and Doppler values are generally much smaller than the signal duration and bandwidth, respectively \cite{Arlery2016, Cui2017,Jing2019, Liu2019,Chen2022,WangF2023}.

Correlation functions can be considered as a special case of AFs in the zero-Doppler cut. In the literature, two important correlation lower bounds are the Welch bound \cite{Welch1974} and the Levenshtein bound \cite{Levenshtein1999, Liu2014, Arlery2019}. These bounds are useful measures of sequence sets operating under static (or quasi-static) wireless environments only. In contrast, the upcoming sixth generation (6G) mobile systems need to 1) deal with highly dynamic environments with high mobility of 1000 km/h (or higher) \cite{Zhang2019, Wang2023} and 2) support efficient design of integrated sensing and communication (ISAC) systems \cite{ITU2023, 3GPPref}. Driven by these two important applications, it is desirable to understand the AF lower bounds in order to measure the optimality of sequence sets in asynchronous mobile channels. Although periodic AF lower bounds are investigated in \cite{Ding2013} and \cite{Ye2022}, a comprehensive study of the aperiodic AF lower bounds is still lacking. There is an aperiodic AF lower bound in \cite{Ye2022} but it is derived by extending the methodologies from the periodic case. Furthermore, the trade-offs between the lower bounds of the aperiodic AAF and CAF are largely untouched, to the best of our knowledge. 

From the construction perspective, while sequences with low aperiodic correlation have been investigated with respect to the Welch bound \cite{Welch1974} and the Pursley-Sarwate criterion \cite{Pursley1976} in \cite{Mow1995,Mow1997, Boothby2017,Gunther2019,Katz2022}, little is known on systematically constructed sequences having low aperiodic AF sidelobes. The asymptotic behavior of Chu sequence pairs \cite{Chu1972} was studied in \cite{Mow1995} and \cite{Mow1997}, showing that the maximum auto-correlation and cross-correlation magnitudes of the primary Chu sequence and its conjugate can asymptotically meet the Welch bound \cite{Welch1974}. This observation demonstrates that the aperiodic Welch bound is asymptotically achievable, highlighting the effectiveness of Chu sequences. However, similar evidence supporting the achievability of the aperiodic AF lower bounds remains absent.

In this paper, compared to the bound in \cite[Theorem 4]{Ye2022}, we introduce tighter aperiodic AF lower bounds for unimodular sequences with certain LAZ. Such bounds are called ``generalized Arlery-Tan-Rabaste-Levenshtein bounds" as we are motivated by some techniques in \cite{Levenshtein1999} and \cite{Arlery2019} focusing on aperiodic correlation lower bounds of polyphase sequence sets. Our core idea is to explore the upper and lower bounds on the Frobenius norm of the weighted auto- and cross-AF matrices by introducing the delay- and Doppler-related weight vectors as well as several useful structural properties of aperiodic AF.

It is shown that, at zero Doppler, the proposed AF lower bounds include not only the Levenshtein bound \cite{Levenshtein1999} and the Arlery-Tan-Rabaste bound \cite{Arlery2019}, but also several other aperiodic correlation bounds, such as the Welch bound \cite{Welch1974}, the Sarwate bound \cite{Sarwate1979}, and the Peng-Fan bound \cite{Peng2004}. Interestingly, our derivation reveals that the optimal Doppler weight vector should be the one that evenly weights the Doppler bins within the LAZ, aligning with the cyclic nature of the aperiodic AF in the Doppler dimension. In contrast, due to the non-cyclic characteristics of the aperiodic AF in the delay dimension, it is found that tighter aperiodic AF bounds can be obtained by certain selected delay weight vectors. 

Additionally, by analyzing the maximum aperiodic AAF and CAF magnitudes of Chu sequences, we show that certain pairs/sets of Chu sequences are asymptotically order-optimal with respect to our proposed AF bounds over certain LAZ. This analysis substantiates the achievability of these bounds. We then demonstrate the tightness of the proposed bounds through asymptotic analysis and numerical examples.

The remainder of this paper is organized as follows. In Section II, we introduce preliminary definitions and describe several properties of aperiodic AF. Section III presents our proposed aperiodic AF lower bounds. In Section IV, we analyze the maximum aperiodic AAF and CAF magnitudes of Chu sequences in certain LAZ. Comparison and tightness analyses of the proposed aperiodic AF lower bounds are presented in Section V. Section VI concludes the paper.

\textit{Notations}: In the following, we use boldface letters to denote matrices and vectors. $(\cdot)^{\ast}$, $(\cdot)^{\text{T}}$, $(\cdot)^{H}$ stand for the conjugate, transpose, and conjugate transpose, respectively. $\left \lceil \cdot \right \rceil $ and $\left \lfloor \cdot \right \rfloor$ represent the round up and round down operations. The imaginary unit is denoted by $j = \sqrt{-1}$. In addition, we define the circulant matrix $\mathbf c(\bm x)$ specified by any vector $\bm{x} = [x_{0},x_{1},\cdots, x_{L-1}]$ of length $L$ as:
\begin{align*}
        \mathbf c(\bm x)=
        \begin{bmatrix}
        x_{0}&x_{1}&\cdots&x_{L-2}&x_{L-1}\\
        x_{L-1}&x_{0}&\cdots&x_{L-3}&x_{L-2}\\
        \vdots&\vdots&\vdots&\vdots&\vdots\\
        x_{1}&x_{2}&\cdots&x_{L-1}&x_{0}
        \end{bmatrix} =
        \begin{bmatrix}
        \mathbf c(\bm x)_{0}\\
        \mathbf c(\bm x)_{1}\\
        \vdots\\
        \mathbf c(\bm x)_{L-1}
        \end{bmatrix},
\end{align*}
where the $i$-th row vector of $\mathbf c(\bm x)$ is denoted as $\mathbf c(\bm x)_{i}$. 

\section{Preliminary}
\subsection{Definitions}
We consider unimodular sequence set $\mathcal{S} = \left \{  \bm x^{m}\right \} _{m=1}^{M}$ with $M$ sequences of length $N$. Each sequence \mbox{$\bm{x}^{m}=\left [x^{m}_{0}, x^{m}_{1}, \cdots, x^{m}_{N-1}\right ]$} satisfies \mbox{$|x^{m}_{t}|^{2}=1$} for \mbox{$t=0,1,\cdots, N-1$}, meaning that the energy of each sequence equals to $N$. 

The discrete aperiodic AF of \mbox{$\bm{x}^{m}, \bm{x}^{m'} \in \mathcal{S}$} at delay shift $\tau$ and Doppler shift $\nu$ is defined as
\begin{align}
    &A_{\bm{x}^{m}, \bm{x}^{m'}}(\tau, \nu) 
    \nonumber\\=&\left\{\begin{array}{ll}\displaystyle\sum_{t=0}^{N-1-\tau} {x}^{m}_{t}{\left ({x}^{m'}_{t+\tau} \right )}^{\ast}e^{\frac {j2\pi\nu t}{N}}, & 0\leq \tau \leq N-1; \\ \displaystyle\sum_{t=0}^{N-1+\tau} {x}^{m}_{t-\tau}{\left ({x}^{m'}_{t} \right )}^{\ast}e^{\frac {j2\pi\nu t}{N}}, & 1-N\leq \tau \leq -1, \end{array}\right.
\end{align}
where $|\tau|,|\nu| \in \mathbb{Z}_N$. When $m\neq m'$, $A_{\bm{x}^{m}, \bm{x}^{m'}}(\tau, \nu)$ is known as the aperiodic CAF; otherwise, aperiodic AAF denoted by $A_{\bm{x}^{m}}(\tau, \nu)$. It is worth noting that the aperiodic correlation can be seen as a special case of the aperiodic AF when $\nu = 0$, which is represented by $R_{\bm{x}^{m}, \bm{x}^{m'}}(\tau)$.

We further define the LAZ in the delay-Doppler plane as 
\begin{equation}
	\bm \Gamma = \left \{ \left ( \tau,\nu \right )|\tau \in \left ( -Z_{x},Z_{x} \right ),\nu \in\left ( -Z_{y},Z_{y} \right )   \right \},  
\end{equation}
where $Z_x, Z_y \in [1, N]$. In practical applications, $Z_x$ and $Z_y$ are determined by the maximum Doppler frequency and the maximum delay \cite{Ye2022, Meng2024}. When $Z_{x}=Z_{y} = N$, $\bm \Gamma$ refers to the global AF region. 



The maximum (non-trivial) aperiodic AAF magnitude ${\theta}_{a}$ of $\mathcal{S}$ and the maximum aperiodic CAF magnitude ${\theta}_{c}$ associated to the LAZ $\bm \Gamma$ are defined by
\begin{subequations}
\begin{align}
{\theta}_{a}^2= & \max_{ \substack{m \in [1,M], ~ (\tau, \nu)\neq (0, 0),\\|\tau|\leq Z_{x}-1,~ |\nu|\leq Z_{y}-1}} | A_{\bm{x}^{m}}(\tau, \nu)|^2,  \\
{\theta}_{c}^2= &  \max_{ \substack{m,~m' \in [1,M],~m\neq m',\\|\tau|\leq Z_{x}-1,~ |\nu|\leq Z_{y}-1}} | A_{\bm{x}^{m}, \bm{x}^{m'}}(\tau, \nu)|^2.
\end{align}
\end{subequations}
The maximum aperiodic (non-trivial) AF magnitude\footnote{Also referred to as peak sidelobe level (PSL) of aperiodic AF.} ${\theta}_{\text{max}}$ of $\mathcal{S}$ associated to $\bm \Gamma$ is defined by
\begin{align}
{\theta}_{\text{max}}^2= \max \left \{ {\theta}_{a}^2, {\theta}_{c}^2 \right \}.
\end{align}
Throughout the paper, such a sequence set $\mathcal{S}$ with the maximum aperiodic AF magnitude ${\theta}_{\text{max}}$ is denoted as $(N, M,{\theta}_{\text{max}}, \bm \Gamma)\text{-}\mathcal{S}$.

\subsection{Properties of Aperiodic AF}
Before presenting the aperiodic AF lower bounds, we demonstrate some inherent structural properties of the aperiodic AF of unimodular sequences through the following lemmas:
\begin{lemma}[Zero delay]
\label{Lm1}
    For any unimodular sequence $\bm x$, its aperiodic AAF satisfies
    \begin{align}
    \label{L1}
        A_{\bm{x}}(0, \nu)  = 0, \quad \quad \forall \,\, \nu \neq 0.
    \end{align}
\end{lemma}
\begin{IEEEproof}
    Let $\bm{x} = [x_{0},x_{1},\cdots, x_{N-1}]$, for any $\nu \neq 0$, we have
    $A_{\bm{x}}(0, \nu)  = \sum_{t=0}^{N-1} x_{t} {\left (x_{t} \right )}^{\ast}e^{\frac{j2\pi\nu t}{N}}=\sum_{t=0}^{N-1}e^{\frac{j2\pi\nu t}{N}}=0$.
\end{IEEEproof}

From the definition of $A_{\bm x^{m},\bm x^{m'}}(\tau, \nu)$, the number of terms in its summation is $N - |\tau|$. By letting $d = N- |\tau|$, $d  \in [1,N]$, one can readily show the following lemma: 
\begin{lemma}
\label{Lm2}
    For any unimodular sequences $\bm x^{m}$ and $\bm x^{m'}$ of length $N$, we have
    \begin{align}
    \label{l2}
        {\left |A_{\bm x^{m},\bm x^{m'}}(\tau, \nu) \right |} \le {d}, \quad \forall \,\, |\tau|, |\nu| \in [0, N-1].
    \end{align}
\end{lemma}

\section{Aperiodic AF Lower Bounds of Unimodular Sequences}
\subsection{Aperiodic AF Lower Bounds Associated to the LAZ}
\label{DDLAZB}
In this subsection, we present our proposed aperiodic AF lower bounds of ${\theta}_{\text{max}}$, ${\theta}_{a}$ and ${\theta}_{c}$ for $\mathcal{S}$ associated to the LAZ $\bm \Gamma$ with \mbox{$1\le Z_{x} \le N$} and $1\le Z_{y} \le N$. First, for a unimodular sequence set $\mathcal{S}$, we define $Z_{y}$ Doppler shifted sequences for each $\bm{x}^{m} \in \mathcal{S}$ as
\begin{align}
\label{10}
    &\widetilde{\bm{x}}^{m, r}= \left [x^{m}_{0}, x^{m}_{1}e^{\frac{j2\pi r}{N}}, \cdots, x^{m}_{N-1}e^{\frac{j2\pi(N-1)r}{N}} \right ],  \nonumber \\&\text{for}\; m = 1,2,\cdots,M\;\text{and}\; r = 0,1,\cdots,Z_y-1.
\end{align}
We define sequences $\bm{x}^{m, r}$ of length $2N-1$ by concatenating $\widetilde{\bm{x}}^{m, r}$ with zero vector $\bm{0}_{1\times N-1}$, as follows:
\begin{align}
\label{F10}
    &\bm{x}^{m, r} =\left [\widetilde{\bm{x}}^{m, r},\bm{0}_{1\times (N-1)}\right ]. 
\end{align}

We define two weight vectors \mbox{$\mathbf w=[w_0, w_1, \cdots,w_{Z_{x}-1}]^{T}$} and \mbox{$\mathbf p=[p_0, p_1, \cdots,p_{Z_y-1}]^{T}$}, for delay and Doppler shifts, respectively. $\mathbf w$ and $\mathbf p$ satisfy the following conditions:
\begin{subequations}
\label{WCS}
    \begin{align}
        &\sum_{i=0}^{Z_x-1}w_i=1, \,\,\, w_i\geq 0, \,\,\,i=0,1, \cdots,Z_x-1, \\
        &\sum_{r=0}^{Z_y-1}p_r=1, \,\,\,\, p_r\geq 0, \,\,\,r=0,1, \cdots,Z_y-1. 
    \end{align}
\end{subequations}


Then, we define the matrix $\mathbf{U}$ of size $M Z_x Z_y \times (2N - 1)$ as follows:
\begin{align}
    \mathbf{U}= \begin{bmatrix}
        {\mathbf{U}^{1}}\\
        {\mathbf{U}^{2}}\\
        \vdots \\
        {\mathbf{U}^{M}}
    \end{bmatrix}, \; \text{with} \; \mathbf{U}^{m}= \begin{bmatrix}
        {\bm u(\bm {x}^{m, 0})}\\
        {\bm u(\bm {x}^{m, 1})}\\
        \vdots\\
        {\bm u(\bm {x}^{m, Z_y-1})}
    \end{bmatrix} \; \\ \text{for} \; m =1,2,\cdots, M, \nonumber
\end{align}
where $\bm u(\bm {x}^{m, r})$ represents the weighted matrix of size $Z_x\times(2N-1)$, constructed from the first $Z_{x}$ row-vectors of the circulant matrix $\mathbf c \left (\bm {x}^{m, r} \right )$ in conjunction with weight vectors $\mathbf w$ and $\mathbf p$. $\bm u(\bm {x}^{m, r})$ is defined as
\begin{align}
\label{uxmr}
    &\bm u(\bm {x}^{m, r}) 
     \!=\! \begin{bmatrix}
    \bm u(\bm {x}^{m, r})_{0}\\
    \bm u(\bm {x}^{m, r})_{1}\\
    \vdots\\
    \bm u(\bm {x}^{m, r})_{Z_x-1}
    \end{bmatrix}  \!=\!  
    \begin{bmatrix}
    \sqrt{p_{r}}\sqrt{w_{0}}\mathbf c(\bm {x}^{m, r})_{0}\\
    \sqrt{p_{r}}\sqrt{w_{1}}\mathbf c(\bm {x}^{m, r})_{1}\\
    \vdots\\
    \sqrt{p_{r}}\sqrt{w_{Z_x-1}}\mathbf c(\bm {x}^{m, r})_{Z_x-1}
    \end{bmatrix}\!.
\end{align}
We set the $(k+1)$-th element of $\bm u(\bm {x}^{m, r})_{i}$ as $ u(\bm {x}^{m, r})_{i, k}$, so we have 
\begin{align*}
    \bm u(\bm {x}^{m, r})_{i}&=[u(\bm {x}^{m, r})_{i, 0}, u(\bm {x}^{m, r})_{i, 1},\cdots,u(\bm {x}^{m, r})_{i, 2N-2}].
\end{align*} 

Note that the following identity is satisfied:
\begin{align}
\label{UU}
  \left \|\mathbf{U}^{H}\mathbf{U} \right\|_{F}^{2}=\left \|\mathbf{U}\mathbf{U}^{H} \right \|_{F}^{2},   
\end{align}
where $\left\|\cdot\right\|_{F}$ represents the Frobenius norm. Specifically, for a matrix $\mathbf{Y} \in \mathbb{C}^{m \times n}$, we have $\|\mathbf{Y}\|_F = \sqrt{\sum_{i=1}^m \sum_{j=1}^n |Y_{ij}|^2}$,
with $Y_{ij}$ denoting the $(i, j)$-th entry of $\mathbf{Y}$.

We first consider the left-hand side of (\ref{UU}).
\begin{lemma}
\label{Lm3}
    $\left \|\mathbf{U}^{H}\mathbf{U} \right \|_{F}^{2}$ can be lower-bounded by 
\begin{align}
    &\left \|\mathbf{U}^{H}\mathbf{U} \right\|_{F}^{2} \geq {M^{2}} \left(N-\sum_{s, t=0}^{Z_x-1}l_{s, t, N}w_{s}w_{t} \right), 
\end{align}
with $l_{s, t, N}=|t-s|$.
\end{lemma}
\begin{IEEEproof}
From the construction of $\mathbf{U}$, we have
\begin{subequations}
    \label{L2P}
    \begin{align}
    \label{L2Pa}
    &\left \|\mathbf{U}^{H}\mathbf{U} \right\|_{F}^{2}
    \nonumber\\= &\sum_{k, k'=0}^{2N-2} \left |\sum_{m=1}^{M}\sum_{r=0}^{Z_y-1}\sum_{i=0}^{Z_x-1} u(\bm {x}^{m, r})_{i, k} \left ( u(\bm {x}^{m, r})_{i, k'} \right )^{\ast}\right|^{2}\\\label{L2Pb}
    \geq &\sum_{k=0}^{2N-2} \left (\sum_{r=0}^{Z_y-1}p_{r}\sum_{m=1}^{M}\sum_{i=0}^{Z_x-1}| c(\bm {x}^{m, 0})_{i, k} |^{2}w_{i} \right )^{2}\\\label{L2Pc}
    = &\sum_{k=0}^{2N-2} \left (\sum_{m=1}^{M}\sum_{i=0}^{Z_x-1}| c(\bm {x}^{m, 0})_{i, k} |^{2}w_{i} \right )^{2} \\\label{L2Pd}
    =&{M^{2}} \left (N-\sum_{s, t=0}^{Z_x-1}l_{s, t, N}w_{s}w_{t} \right ),
    \end{align}
\end{subequations}
where the inequality (\ref{L2Pb}) is obtained by removing all terms $k \neq k'$ in (\ref{L2Pa}). Equation (\ref{L2Pc}) is derived based on the condition (\ref{WCS}b) of the weight vector $\mathbf p$. The proof of (\ref{L2Pd}) is similar to \cite[Appendix B]{Arlery2019} and hence omitted here.
\end{IEEEproof}

\begin{strip}
\rule{\textwidth}{0.4pt} 
    \emph{Example:} Set $N=3$, $M=2$, $Z_x=3$, and $Z_y=2$. The matrix $\mathbf{U}$ results in a size of $12 \times 5$:
\begin{align*}
    \mathbf{U}={\begin{bmatrix}
        \sqrt{p_{0}}\sqrt{w_{0}}x_{0}^{1}&\sqrt{p_{0}}\sqrt{w_{0}}x_{1}^{1}&\sqrt{p_{0}}\sqrt{w_{0}}x_{2}^{1}&0&0\\
        0&\sqrt{p_{0}}\sqrt{w_{1}}x_{0}^{1}&\sqrt{p_{0}}\sqrt{w_{1}}x_{1}^{1}&\sqrt{p_{0}}\sqrt{w_{1}}x_{2}^{1}&0\\
        0&0&\sqrt{p_{0}}\sqrt{w_{2}}x_{0}^{1}&\sqrt{p_{0}}\sqrt{w_{2}}x_{1}^{1}&\sqrt{p_{0}}\sqrt{w_{2}}x_{2}^{1}\\
        \sqrt{p_{1}}\sqrt{w_{0}}x_{0}^{1}&\sqrt{p_{1}}\sqrt{w_{0}}x_{1}^{1}e^{\frac{j2\pi}{N}}&\sqrt{p_{1}}\sqrt{w_{0}}x_{2}^{1}e^{\frac{j4\pi}{N}}&0&0\\
        0&\sqrt{p_{1}}\sqrt{w_{1}}x_{0}^{1}&\sqrt{p_{1}}\sqrt{w_{1}}x_{1}^{1}e^{\frac{j2\pi}{N}}&\sqrt{p_{1}}\sqrt{w_{1}}x_{2}^{1}e^{\frac{j4\pi}{N}}&0\\
        0&0&\sqrt{p_{1}}\sqrt{w_{2}}x_{0}^{1}&\sqrt{p_{1}}\sqrt{w_{2}}x_{1}^{1}e^{\frac{j2\pi}{N}}&\sqrt{p_{1}}\sqrt{w_{2}}x_{2}^{1}e^{\frac{j4\pi}{N}}\\
        \sqrt{p_{0}}\sqrt{w_{0}}x_{0}^{2}&\sqrt{p_{0}}\sqrt{w_{0}}x_{1}^{2}&\sqrt{p_{0}}\sqrt{w_{0}}x_{2}^{2}&0&0\\
        \vdots&\vdots&\vdots&\vdots&\vdots\\
        0&0&\sqrt{p_{1}}\sqrt{w_{2}}x_{0}^{2}&\sqrt{p_{1}}\sqrt{w_{2}}x_{1}^{2}e^{\frac{j2\pi}{N}}&\sqrt{p_{1}}\sqrt{w_{2}}x_{2}^{2}e^{\frac{j4\pi}{N}}
        \end{bmatrix}}_{12 \times 5}.
\end{align*}
\end{strip}

Now we consider the right-hand side of (\ref{UU}).
\begin{lemma}
\label{Lm4}
$\left \|\mathbf{U}\mathbf{U}^{H}\right \|_{F}^{2}$ can be upper-bounded by
\begin{align}
\label{L4sep}
    ||\mathbf{U}\mathbf{U}^{H}||_{F}^{2}\leq & \theta_{c}^{2} M(M-1)\left(1-\sum_{d=E}^{D}\sum_{\substack{s, t=0 \\l_{s, t, N}=N-d} }^{Z_x-1}w_{s}w_{t} \right )\nonumber\\& + \theta_{a}^{2} M\left(1-\! \sum_{i=0}^{Z_{x}-1} w_{i}^{2}-\!\sum_{d=E}^{D}\!\sum_{\substack{s, t=0 \\l_{s, t, N}=N-d} }^{Z_x-1}\!w_{s}w_{t} \right ) \nonumber\\
& + MN^{2}\sum_{r=0}^{Z_{y}-1}p_{r}^{2}\sum_{i=0}^{Z_{x}-1}  w_{i}^{2} \nonumber\\&+ M^{2}\sum_{d=E}^{D}d^2 \sum_{\substack{s, t=0 \\l_{s, t, N}=N-d} }^{Z_x-1}w_{s}w_{t},
\end{align}
and
    \begin{align}
    \label{L4}
        ||\mathbf{U}\mathbf{U}^{H}||_{F}^{2} \leq& M^{2}{\theta}_{\text{max}}^{2}+ M \left (N^2\sum_{r=0}^{Z_y-1}p_{r}^{2}- {\theta}_{\text{max}}^{2}\right )\sum_{i=0}^{Z_x-1} w_{i}^{2}\nonumber \\
        & -\sum_{d=E}^{D}\left (M^{2}\left ({\theta}_{\text{max}}^{2}-{d^{2}}\right )\sum_{\substack{s, t=0\\l_{s, t, N}=N-d} }^{Z_x-1}w_{s}w_{t} \right ),
    \end{align}
where $D \in [0,N-1]$ represents the number of last delays taken into consideration and $E = N-Z_x+1$.
\end{lemma}
\begin{IEEEproof}
    Based on \textit{Lemma~\ref{Lm1}} and \textit{Lemma~\ref{Lm2}}, we have
\begin{align}
\label{L4P}
&||\mathbf{U}\mathbf{U}^{H}||_{F}^{2}\nonumber\\=&\sum_{m,m'=1}^{M}\sum_{r,r'=0}^{Z_{y}-1}\sum_{i,i'=0}^{Z_{x}-1} |A_{\bm{x}^{m}, \bm{x}^{m'}}(i-i', r-r')|^{2}p_{r}p_{r'} w_{i}w_{i'} \nonumber\\
\leq&\sum_{\substack{m,m'=1\\m\neq m'}}^{M}\sum_{r,r'=0}^{Z_{y}-1}\sum_{i,i'=0}^{Z_{x}-1} {\theta}_{c}^{2} p_{r}p_{r'}w_{i}w_{i'}\nonumber\\&+\sum_{m=1}^{M}\sum_{r,r'=0}^{Z_{y}-1}\sum_{i,i'=0}^{Z_{x}-1} {\theta}_{a}^{2} p_{r}p_{r'}w_{i}w_{i'} \nonumber\\
& + \!M\!\left(N^{2}\! - \!{\theta}_{a}^{2} \right)\! \sum_{r=0}^{Z_{y}-1}\!p_{r}^{2}\!\sum_{i=0}^{Z_{x}-1}  \!w_{i}^{2}\! -\!  \sum_{m=1}^{M}\sum_{\substack{r,r'=1\\r\neq r'}}^{Z_{y}-1}\!\sum_{i=0}^{Z_{x}-1} \!{\theta}_{a}^{2} p_{r}p_{r'} w_{i}^{2} \nonumber\\
& - M (M-1) \sum_{d=E}^{D} \left ( {\theta}_{c}^{2} - d^{2}\right ) \!\!\!\sum_{\substack{s, t=0\\l_{s, t, N}=N-d} }^{Z_x-1} \!\!\! w_{s}w_{t} 
\nonumber\\&-M \sum_{d=E}^{D} \left( {\theta}_{a}^{2} - d^{2} \right) \!\!\!\sum_{\substack{s, t=0\\l_{s, t, N}=N-d} }^{Z_x-1} \!\!\! w_{s}w_{t}.
\end{align}
By substituting (\ref{WCS}) into (\ref{L4P}), we conclude (\ref{L4sep}) and (\ref{L4}).
\end{IEEEproof}

Based on (\ref{L4P}), when the following property is satisfied:
\begin{align}
\label{CD}
    \exists\,\,(w_s,w_t)\,\, \text{s.t.} \sum_{\substack{s, t=0\\l_{s, t, N}=N-d} }^{Z_x-1}w_{s}w_{t} \neq 0, \quad {\theta}_{\text{max}}^{2} > {d^2}, 
\end{align}
there exists a positive integer $D>E-1$ such that (\ref{L4}) is tighter than the case with $D=0$. A similar property for (\ref{L4sep}) can also be easily derived.

Based on the weighting condition (\ref{WCS}b) and by utilizing the Cauchy-Schwarz inequality, we can determine that the minimum value of $\sum_{r=0}^{N-1}p_{r}^{2}$ in (\ref{L4}) is $\frac{1}{Z_y}$. This minimum value is achieved if and only if 
\begin{align}
    \label{optweightp}
    \mathbf{p} = \hat{\mathbf{p}} = \left [\frac{1}{Z_y}, \frac{1}{Z_y},\cdots,\frac{1}{Z_y} \right].
\end{align}
Hence, $\hat{\mathbf{p}}$ is the optimal choice of weight vector ${\mathbf{p}}$ for \textit{Lemma \ref{Lm4}}, as it yields the tightest upper bound of $\left \|\mathbf{U}\mathbf{U}^{H}\right \|_{F}^{2}$.

\begin{remark}
    It is interesting to note that since \textit{Lemma~\ref{Lm3}} is independent of the weight vector $\mathbf{p}$, while the optimal choice of $\mathbf{p}$ for \textit{Lemma~\ref{Lm4}} is $\hat{\mathbf{p}}$ in (\ref{optweightp}), therefore $\hat{\mathbf{p}}$ is also the optimal Doppler weight vector for the desired aperiodic AF lower bounds.
\end{remark}

Then, from the above lemmas and this optimal weight vector $\hat{\mathbf{p}}$, we can deduce the aperiodic AF lower bounds with respect to the LAZ. We first define matrices $\bm J_{Z_x}^{d}$ and $\bm L_{Z_x}$ such that:
\begin{align}
  \sum_{\substack{s, t=0\\l_{s, t, N}=N-d}}^{Z_x-1}w_{s}w_{t}&=\mathbf{w}^{T}\bm J_{Z_x}^{d}\mathbf{w}, \nonumber \\ \sum_{s, t=0}^{Z_x-1}l_{s, t, N}w_{s}w_{t}&=\mathbf{w}^{T}\bm{L}_{Z_x}\mathbf{w}. \nonumber
\end{align}

\begin{theorem} [Aperiodic AF lower bounds associated to an LAZ]
\label{T1fff}
    For any weight vector $\mathbf w$ satisfying the weighting condition (\ref{WCS}a), aperiodic AF lower bounds associated to any LAZ $\bm \Gamma$ ($1< Z_{x} \le N$ and $1\le Z_{y} \le N$) for any sequence set $(N, M,{\theta}_{\text{max}}, \bm \Gamma)\text{-}\mathcal{S}$ is given by:
    \begin{subequations}
        \label{T2}
        \begin{align}
    &{\theta}_{c}^{2} (M-1)\left(1-\mathbf{w}^{T}\sum_{d=E}^{D}\bm J^{d}_{Z_{x}}\mathbf{w} \right)\nonumber\\+&{\theta}_{a}^{2} \left (1-\mathbf{w}^{T}\left(\bm I_{Z_{x}}+\sum_{d=E}^{D}\bm J^{d}_{Z_{x}}\right)\mathbf{w}\right ) \nonumber\\
\geq& M\left(N-Q\left (\mathbf{w}, \frac{N^{2}}{MZ_{y}},\sum_{d=E}^{D}d^{2}{\bm J}^{d}_{Z_{x}} \right )\right),\\
        {\theta}_{\text{max}}^{2}\geq& {N}-\frac{{\mathcal{Q}}\left (\mathbf w, \frac{N(N-Z_{y})}{MZ_{y}}, \sum\limits_{d=E}^{D}(d^{2}-N){\bm J}^{d}_{Z_x}\right ) }{ 1-\mathbf{w}^{T} \left (\frac{1}{M}{\bm I}_{Z_x}+\sum\limits_{d=E}^{D}{\bm J}^{d}_{Z_x} \right)\mathbf{w}},
    \label{T2b}
    \end{align}
    \end{subequations}
    where ${\mathcal{Q}}(\mathbf w, \eta, \bm{B})= \mathbf{w}^{T}(\eta \bm{I}_{Z_x}+\bm{B}+\bm{L}_{Z_x})\mathbf{w}$, $E = N-Z_x+1$ and $D \in [0,N-1]$.
\end{theorem}
\begin{IEEEproof}
   Based on equation (\ref{UU}), substituting the optimal weight vector $\hat{\mathbf{p}}$ from (\ref{optweightp}) into \textit{Lemma~\ref{Lm4}} and combining it with \textit{Lemma~\ref{Lm3}} yields (\ref{T2}).
\end{IEEEproof}

As special cases, the correlation lower bound and AF lower bound for global AF can be derived from \textit{Theorem \ref{T1fff}}.

\begin{remark}
When $Z_y = 1$ and $D = 0$, the proposed bounds in (\ref{T2}) reduce to the Peng-Fan bounds \cite[Theorem 2]{Peng2004} for aperiodic correlation in the low correlation zone. This correlation bound is equivalent to the AF bound for $Z_y > 1$ derived from \textit{Lemma \ref{Lm3}} and \textit{Lemma \ref{Lm4}} using the non-optimal weight vector $\mathbf{p}=[1, 0, \cdots, 0]^T$. Therefore, for cases with non-zero Doppler shifts, the proposed bound (\ref{T2}) is strictly tighter than the Peng-Fan bound \cite[Theorem 2]{Peng2004}, owing to the optimal weight vector $\hat{\mathbf{p}}$.
\end{remark}

\begin{corollary} [Aperiodic AF lower bound for global AF]
\label{Pp1}
    For any set $\mathcal{S}$ of $M$ unimodular sequences of length $N$, the global aperiodic AF lower bound (i.e., $Z_{x}=Z_{y} = N$) is given by:
    \begin{align}
        \label{GAF}
        &{{\theta}^2_{\text{max}}} \ge \left\{\begin{matrix}
        {N-1}, & M=1;\\
        {N}, & M>1.
        \end{matrix}\right.
    \end{align}
For any LAZ $\bm \Gamma$ with $1<Z_{x}\le N$ and $Z_{y}=N$, the bound in (\ref{GAF}) still holds.
\end{corollary}

\begin{IEEEproof}
For any delay $\tau$ in aperiodic scenario, define $\bm x_{t+\tau}^{m'}=0$ for $|t+\tau|\geq N$, we have
    \begin{align}
    \label{L1P}
       &\sum_{\nu = 0}^{N-1} {\left |A_{\bm{x}^{m}, \bm{x}^{m'}}(\tau, \nu) \right |}^2 
       \nonumber\\= & \sum_{t=0}^{N-1}\sum_{s=0}^{N-1} x_{t}^{m} {\left ( x_{t+\tau}^{m'} \right )}^{\ast}  {\left (x_{s}^{m} \right )}^{\ast} x_{s+\tau}^{m'} \sum_{\nu=0}^{N-1}e^{\frac{j2\pi\nu (t-s)}{N}} \nonumber \\
       =& N\sum_{t=0}^{N-1} {\left |x_{t}^{m} \right |}^2{\left |x_{t+\tau}^{m'} \right |}^2 =N\left ({N-|\tau|} \right ).
    \end{align}
For $Z_{x}=Z_{y} = N$, according to (\ref{L1P}) and \textit{Lemma \ref{Lm1}}, we have
    \begin{subequations}
        \begin{align}
         {\theta}_{c}^2
        &\ge \max_{ \substack{m,~m' \in [1,M],~m\neq m',\\|\tau|\leq Z_{x}-1}}\left\{\frac{\sum_{\nu = 0}^{N-1} {\left |A_{\bm{x}^{m}, \bm{x}^{m'}}(\tau, \nu) \right |}^2}{N}\right \} 
        \nonumber \\ &\ge  \max_{ |\tau|\leq Z_{x}-1} \left \{ N-|\tau| \right \} = N; \label{28caf}\\
        {\theta}_{a}^2
        &\ge \max_{ \substack{m \in [1,M],\\ 1 \leq |\tau|\leq Z_{x}-1}}\left\{\frac{\sum_{\nu = 0}^{N-1} {\left |A_{\bm{x}^{m}, \bm{x}^{m'}}(\tau, \nu) \right |}^2}{N}\right \} 
        \nonumber \\ &\ge  \max_{ 1\leq |\tau|\leq Z_{x}-1} \left \{ N-|\tau| \right \} = N-1,
        \label{28aaf}
    \end{align}
    \end{subequations}
    For $M>1$, the global AF lower bound of ${\theta^2_{\text{max}}}$ in (\ref{GAF}) is determined by the CAF bound (\ref{28caf}); for $M=1$, it is determined by the AAF bound (\ref{28aaf}).
\end{IEEEproof}

\begin{remark}
    The lower bound for global AF in (\ref{GAF}) is also a special case of the proposed bound (\ref{T2b}) for delay-Doppler LAZ. When $Z_y = N$ and $M \ge 2$, the proposed bound (\ref{T2b}) is equal to (\ref{GAF}) by choosing weight vector $\mathbf{w} = [1, 0, \cdots, 0]^{T}$. When $Z_y = N$ and $M = 1$, the proposed bound (\ref{T2b}) is equal to (\ref{GAF}) with weight vector $\mathbf{w} = [0.5, 0.5, 0, \cdots, 0]^{T}$.
\end{remark}

Next, we investigate different weight vectors $\mathbf w$ for the bound (\ref{T2}). We first consider (\ref{T2}) with $D=0$.

\subsubsection{Weight vector A ($\mathbf w^{A,q}$)}
An intuitive weight vector, denoted as $\mathbf w^{A,q}$, can be defined as
\begin{align}
    w_{i}^{A,q}=\left\{\begin{array}{ll}\frac{1}{q}, & 0\leq i \leq q-1; \\ 0, & q \leq i \leq Z_x-1,\end{array}\right.
\end{align}
where $1\leq q \leq Z_x$.

By appropriately selecting $q$, we obtain:

\begin{corollary}
\label{COROA}
For any LAZ $\bm \Gamma$ satisfying $Z_{x} > \sqrt{\frac{3N^{2}}{MZ_{y}}}$ with the condition that $MZ_{y}\geq 3$, we have
\begin{subequations}
    \label{C3}
    \begin{align}
    &{\theta}_{c}^{2} (M-1)+{\theta}_{a}^{2}\left (1-\frac{\sqrt{MZ_{y}}}{\sqrt{3}N} \right )
\geq MN\frac{\sqrt{3MZ_{y}}-2}{\sqrt{3MZ_{y}}},\\
         &{\theta}_{\text{max}}^{2}  \geq N-\frac{2N}{\sqrt{3MZ_y}}.
    \end{align}
\end{subequations}
\end{corollary}

\begin{IEEEproof}
See Appendix~{\ref{appA}}.
\end{IEEEproof}

\subsubsection{Weight vector B ($\mathbf w^{B,q}$)}
We derive the weight vector $\mathbf w^{B,q}$ using a quadratic minimization approach, similar to \cite[Lemma 2]{Levenshtein1999} and \cite{Liu2017}. The resulting weight vector is expressed as:
\begin{align}
\label{WV1}
    w_{i}^{B,q}=\left\{\begin{array}{ll}
        \frac{\sin \frac{\gamma}{2}}{\sin \frac{q\gamma}{2}}\sin \left (\gamma_{0}+i\gamma \right ), & 0 \le i \le q-1; \\ 
        0, & q \le i \le Z_x-1,
    \end{array}\right.
\end{align}
where \(q\) is an even positive integer satisfying \(q\gamma \leq \pi+\gamma\), and \(\gamma_0= \frac{\pi-q\gamma+\gamma}{2}\). Here, \(\gamma=\arccos \left (1-\frac{MZ_{y}}{N^{2}} \right )\) and \(MZ_{y} \leq N^{2}\).

By substituting $\mathbf w^{B,q}$ and \(D=0\) into the bound (\ref{T2}), and selecting \(q = \left \lfloor \frac{\pi}{\gamma} \right \rfloor + 1\), we obtain the following result.

\begin{corollary}
\label{COROB}
    For any LAZ $\bm \Gamma$ satisfying $Z_x > \frac{\pi}{\gamma}$ and $5 \leq MZ_{y} \leq N^2$, we have
    \begin{align}
    \label{C5}
        {\theta}_{\text{max}}^{2} &\geq \frac{{\theta}_{c}^{2} (M-1)+{\theta}_{a}^{2}}{M} \nonumber \\&\geq N-\left \lceil\frac{\pi N}{\sqrt{8MZ_{y}}} \right \rceil.
    \end{align}
\end{corollary}

Based on these weight vectors above, we consider other values of $D$ for (\ref{T2}). We denote the optimal $D$ as $D_{\text{opt}}$, which makes (\ref{T2b}) the tightest under a given weight vector among different values of $D$. If (\ref{CD}) is satisfied for a weight vector $\mathbf w$, then there exists \mbox{$D_{\text{opt}}>E-1$}. Let (\ref{T2b}) with $D = 0$ be denoted as $\text{AFB}_{\text{ref}}$. Then, from \textit{Lemma~\ref{Lm2}} and (\ref{CD}), we have $D_{\text{opt}} \approx \left \lfloor \sqrt{\text{AFB}_{\text{ref}}} \right \rfloor$. Conversely, consideration of $D_{\text{opt}}$ is only necessary when \mbox{$Z_x \geq N-\left \lfloor \sqrt{\text{AFB}_{\text{ref}}} \right \rfloor$}. For weight vectors such as $\mathbf{w}^{A,q}$ and $\mathbf{w}^{B,q}$ which have only the first $q$ elements non-zero, this condition becomes $q \geq N-\left \lfloor \sqrt{\text{AFB}_{\text{ref}}} \right \rfloor$.

\subsection{LAZ with $Z_{x}=N$ Case}
Aperiodic AF lower bound associated to the LAZ $\bm \Gamma$ with $Z_{x}=N$ and $1\le Z_{y} \le N$ can be considered as a special case of (\ref{T2}). Nevertheless, a generalization of the weight vector $\mathbf{w}$ for this $Z_{x}=N$ case can be achieved. Essentially, the bound (\ref{T2}) leverages the first $Z_{x}$ row-vectors of the circulant matrix $\mathbf c \left (\bm {x}^{m, r} \right )$ in (\ref{uxmr}). Due to the cyclic property of $\mathbf c \left (\bm {x}^{m, r} \right )$, when $Z_{x}=N$, $\bm u(\bm {x}^{m, r})$ in (\ref{uxmr}) can incorporate all the $2N-1$ row-vectors of $\mathbf c \left (\bm {x}^{m, r} \right )$. In other words, when the maximum delay of interest is $N-1$, the absolute value of the delay between any two row-vectors in $\mathbf c \left (\bm {x}^{m, r} \right )$ is less than or equal to $N-1$. Thus, the weighting condition (\ref{WCS}a) can be replaced by:
\begin{align}
\label{WCS2}
    &\sum_{i=0}^{2N-2}w_i=1, \,\,\, w_i\geq 0, \,\,\,i=0,1, \cdots,2N-2.  
\end{align}
Following the same steps as in (\ref{10}) to (\ref{T2}), replacing $l_{s, t, N} = |t-s|$ with the extended definition $l'_{s, t, N} = \min \left \{|t-s|, 2N-1-|t-s| \right \}$, generalized aperiodic AF lower bounds for the LAZ $\bm{\Gamma}$ with $Z_{x}=N$ can be derived as follows:
\begin{theorem}[Aperiodic AF lower bounds for LAZ with $Z_{x}=N$]
    For any weight vector $\mathbf w$ satisfying the weighting condition (\ref{WCS2}), and any LAZ $\bm \Gamma$ with $Z_{x}=N$ and $1\le Z_{y} \le N$, the aperiodic AF lower bound for any sequence set $(N, M,{\theta}_{\text{max}}, \bm \Gamma)\text{-}\mathcal{S}$ is given by:
    \begin{subequations}
        \label{T1F}
         \begin{align}
        &{\theta}_{c}^{2} (M-1)\left(1-\mathbf{w}^{T}\sum_{d=1}^{D}\bm J^{d}_{2N-1}\mathbf{w} \right)\nonumber \\ +&{\theta}_{a}^{2} \left (1-\mathbf{w}^{T}\left(\bm I_{2N-1}+\sum_{d=1}^{D}\bm J^{d}_{2N-1}\right)\mathbf{w}\right ) \nonumber\\
\geq& M\left(N-Q\left (\mathbf{w}, \frac{N^{2}}{MZ_{y}},\sum_{d=1}^{D}d^{2}{\bm J}^{d}_{2N-1} \right )\right),\\
        {\theta}_{\text{max}}^{2} \geq&  {N} -\frac{{\mathcal{Q}} \left (\mathbf w, \frac{N(N-Z_{y})}{MZ_{y}}, \sum\limits_{d=1}^{D}(d^{2}-N)\bm J_{2N-1}^{d} \right ) }{1-\mathbf{w}^{T} \left (\frac{1}{M}{\bm I}_{2N-1}+\sum\limits_{d=1}^{D}\bm J_{2N-1}^{d} \right)\mathbf{w}},
        \label{T1}
    \end{align}
    \end{subequations}
    where ${\mathcal{Q}}(\mathbf w, \eta, \bm{B})= \mathbf{w}^{T}(\eta\bm{I}_{2N-1}+\bm{B}+\bm{L}_{2N-1})\mathbf{w}$ and \mbox{$D \in [0,N-1]$}.
\end{theorem}

\begin{remark}
    \begin{itemize}
        \item When $Z_y = 1$, the proposed bound (\ref{T1}) reduces to the Arlery-Tan-Rabaste bound \cite[Theorem 2]{Arlery2019}, which corresponds to the Levenshtein bound when $D = 0$ \cite[Theorem 1]{Levenshtein1999}. Again, these correlation bounds are equivalent to the AF bounds for $Z_y > 1$ obtained using the non-optimal weight vector $\mathbf{p}=[1, 0, \cdots, 0]^T$, so the proposed bound (\ref{T1}) is strictly tighter than them for $Z_y > 1$.
        \item Compared to the original weighting condition (\ref{WCS}a) with $Z_{x}=N$, weighting condition (\ref{WCS2}) provides enhanced flexibility in selecting weight vectors for LAZ with $Z_{x}=N$ case. For a weight vector $\mathbf{w}$ with non-zero values present only in the first $N$ elements, the bounds (\ref{T1F}) and (\ref{T2}) are equivalent.
    \end{itemize}
\end{remark}

\subsubsection{Weight vector C ($\mathbf w^C$)}
We demonstrate the significance of the new weighting condition (\ref{WCS2}) by using the most straightforward weight vector $\mathbf w^C$, which is defined as
\begin{align}
    w_{i}^C = \frac{1}{2N-1}, \quad i = 0,1,\cdots,2N-2.
\end{align}

\begin{corollary}
    For any LAZ $\bm \Gamma$ with $Z_{x}=N$ and $1\le Z_{y} \le N$, we have
    \begin{subequations}
    \label{C1F}
            \begin{align}
            \label{C1a}
        &{\theta}_{c}^{2} (M-1)+{\theta}_{a}^{2}\frac{2N}{2N-1} 
\geq N^2\frac{MZ_{y}-1}{(2N-1)Z_{y}},\\
         &{\theta}_{\text{max}}^{2}  \geq N^2\frac{MZ_{y}-1}{M(2N-1)Z_{y}-Z_{y}}.
        \label{C1}
    \end{align} 
    \end{subequations}
\end{corollary}

\begin{IEEEproof}
The proof of (\ref{C1F}) is similar to that of \textit{Corollary \ref{COROA}} by substituting $\mathbf{w}^{C}$ and $D = 0$ into (\ref{T1}), note that $\sum_{s, t=0}^{2N-2}l'_{s, t, N}w_{s}w_{t} = \frac{N(N-1)}{2N-1}$ for $\mathbf{w}^{C}$.
\end{IEEEproof}

\begin{remark}
    When $Z_y=1$, bound (\ref{C1a}) reduces to the Sarwate bound for auto- and cross-correlation \cite{Sarwate1979}, bound (\ref{C1}) reduces to the Welch bound \cite{Welch1974} for aperiodic correlation.
\end{remark}

Based on the following analysis, $\mathbf{w}^{C}$ is the optimal weight vector for bound (\ref{T1}) with $D=0$ in certain scenarios. Specifically, for the quadratic function ${\mathcal{Q}}(\mathbf w, \frac{N(N-Z_{y})}{MZ_{y}}, \bm 0_{2N-1})$ in (\ref{T1}), when the following condition of $M$, $N$ and $Z_y$ is satisfied \cite{Berlekamp1968}:
\begin{align*}
\label{lkc}
\forall \, k \in[1,2N-1] \,\,\text{s.t.} \,\,\lambda_{k}\! =\!\frac{N(N-Z_{y})}{MZ_{y}}\! - \!\frac{1}{4\sin^{2}\frac{\pi}{2N-1}} \ge 0,
\end{align*}
weight vector $\mathbf w^{C}$ minimizes ${\mathcal{Q}}(\mathbf w,  \frac{N(N-Z_{y})}{MZ_{y}}, \bm 0_{2N-1})$. Additionally, $\mathbf w^{C}$ also minimizes $\mathbf{w}^{T}\mathbf{w}$ in (\ref{T1}), and consequently, it maximizes (\ref{T1}) in the case when $D=0$. Based on these observations, the following remark can be made.
\begin{remark}
\label{R2}
The bound in (\ref{C1}) cannot be further improved by substituting any other weight vector $\mathbf w$ into (\ref{T1}) with $D = 0$ in the following scenarios:
\begin{itemize}
    \item $N \ge 2$ with $M = Z_{y} = 1$ or $M = 2$ and $Z_{y} = 1$;
    \item $N \ge 7$ with $M = 1$ and $Z_{y} = 2$.
\end{itemize}
\end{remark}


Again, for a given weight vector $\mathbf w$ that satisfies (\ref{WCS2}) and
\begin{align}
    \exists\,\,(w_s,w_t)\,\, \text{s.t.} \sum_{\substack{s, t=0\\l'_{s, t, N}=N-d} }^{2N-2}w_{s}w_{t} \neq 0, \quad {\theta}_{\text{max}}^{2} > {d^2}, 
\end{align}
we can conclude that there exist $D_{\text{opt}}>0$. Let (\ref{T1}) with $D = 0$ be denoted as ${\widetilde{\text{AFB}}_{\text{ref}}}$, we have $D_{\text{opt}} \approx \left \lfloor \sqrt{\widetilde{\text{AFB}}_{\text{ref}}} \right \rfloor$. 

\section{Asymptotically Order-Optimal Sequence Set for AF in the LAZ}
\label{Chusection}
According to \textit{Corollary 1-4}, $\theta_{\text{max}}$ should be at least $\mathcal{O}(\sqrt{N})$, when the order of the size of the LAZ (i.e. the order of $Z_{x}Z_{y}$) is no less than $\mathcal{O}(N)$ (see Table I for more examples). Therefore, a sequence set with $\theta_{\text{max}}$ asymptotically equal to $\mathcal{O}(\sqrt{N})$ is said to be ``order-optimal" in this paper. 

To find such order-optimal sequence sets, we consider Chu sequences \cite{Chu1972}. A Chu sequence set can be defined as $\mathcal{A} = \left \{ \mathbf{s}^{m}\right \} _{m=1}^{M}$, with each $\mathbf{s}^{m}=\left [{s}^{a_m}_{0}, {s}^{a_m}_{1}, \cdots,{s}^{a_m}_{N-1}\right ]$ is unimodular sequences of length $N$ given by
\begin{align}
    {s}^{a_m}_t = e^{\frac{\pi j a_m t^2}{N}},
\end{align}
where $a_m$ are distinct integers for different $m$ with $|a_m| \in [1, N-1]$ \cite{Gunther2019}. Note that \cite{Chu1972} used a slightly different definition when $N$ is odd.

We first analyze the maximum AAF magnitude in certain LAZ for Chu sequences. Consider a Chu sequence $\mathbf{s}=\left [{s}^{a}_{0},{s}^{a}_{1}, \cdots, {s}^{a}_{N-1}\right ]$, we derive the following conclusion.

\begin{theorem}[Maximum AAF magnitude in the LAZ for Chu sequences]
\label{T3ss}
    If $|a|>1$, for $-|a|< \nu <|a|$ and $|\tau| \leq\frac{\beta N}{|a|}-1$, with $\frac{1}{2}<\beta<1$, $N\geq5|a|$, we have
    \begin{align}
        \label{TC3}
        \lim_{N \to +\infty}  \frac{\max_{(\tau,\nu)} \left |A_{\mathbf{s}}(\tau, \nu) \right |}{\sqrt{N}} & =\frac{0.4802}{\sqrt{|a|}}.
    \end{align}
\end{theorem}

\begin{IEEEproof}
   See Appendix~\ref{appB}.
\end{IEEEproof}

This indicates that the maximum (non-trivial) aperiodic AAF magnitude of the Chu sequence in an LAZ satisfying the conditions in \textit{Theorem \ref{T3ss}} is asymptotically in the order of $\mathcal{O}(\sqrt{N})$.

We then analyze the maximum CAF magnitude in the LAZ for Chu sequences. Consider two Chu sequence $\mathbf{s}^{1}=\left [{s}^{a_1}_{0}, {s}^{a_1}_{1}, \cdots, {s}^{a_1}_{N-1}\right ]$ and $\mathbf{s}^{2}=\left [{s}^{a_2}_{0}, {s}^{a_2}_{1}, \cdots, {s}^{a_2}_{N-1}\right ]$ with $a_{1} > a_{2}$. We need the following lemma on the bounding method of Van Der Corput \cite{ferenczi2018ergodic}.

\begin{lemma}[{\cite[Theorem 8.15]{ferenczi2018ergodic}}]
    \label{lvdc}
Let $\mathbb{I}$ be an interval of $\mathbb{R}$ containing $\xi$ integers ($\xi\geq0$). Let $f : \mathbb{I} \rightarrow \mathbb{R}$ be a twice continuously differentiable function on $\mathbb{I}$. If there exists a real number $\rho > 0$ and a real number $\alpha\geq 1$ such that $\forall n\in \mathbb{I}, \,\, \rho  \leq|f^{''}(n)|\leq \alpha \rho$, the exponential sum of $e^{j2\pi f(n)}$ satisfies
\begin{align}
\label{lem5}
    \left|\sum_{n\in \mathbb{I}} e^{j2\pi f(n)}\right |\leq 3\alpha \xi \sqrt{\rho}+\frac{6}{\sqrt{\rho}}.
\end{align}
\end{lemma}

\begin{theorem}[Maximum CAF magnitude in the LAZ for Chu sequences]
    \label{T4Css}
    For a Chu sequence pair $\mathbf{s}^{1}$ and $\mathbf{s}^{2}$ with $a_1>a_2$, we have
    \begin{align}
    \label{T4}
        &{\left |A_{\mathbf{s}^{1}, \mathbf{s}^{2}}(\tau, \nu) \right |}\nonumber\\\leq& 3\left (\sqrt{a_{1}-a_{2}}+\frac{2}{\sqrt{a_{1}-a_{2}}} \right )\sqrt{N} -\frac{3|\tau|\sqrt{a_{1}-a_{2}}}{\sqrt{N}}.
    \end{align}
\end{theorem}

\begin{IEEEproof}   
For $\tau>0$, we have
\begin{align}
{\left |A_{\mathbf{s}^{1}, \mathbf{s}^{2}}(\tau, \nu) \right |} & =\left|\sum_{n=0}^{N-1-\tau} s^{1}_{n}s_{n+\tau}^{2\ast}e^{\frac{j2\pi\nu n}{N}}\right| \nonumber\\
&=\left |\sum_{n=0}^{N-1-\tau} e^{j\pi\frac{(a_{1}-a_{2})n^{2}+2(\nu-a_{2}\tau)n-a_{2}\tau^{2}}{N}}\right |\nonumber\\
& =\left|\sum_{n=0}^{N-1-\tau} e^{j\pi\frac{(a_{1}-a_{2})n^{2}+2(\nu-a_{2}\tau)n}{N}} \right|.
\end{align}
Similarly, we have ${\left |A_{\mathbf{s}^{1}, \mathbf{s}^{2}}(\tau, \nu) \right |}=\left|\sum_{n=0}^{N-1-\tau} e^{j\pi\frac{(a_{1}-a_{2})n^{2}+2(\nu-a_{1}\tau)n}{N}} \right|$, for $\tau<0$.

According to \textit{Lemma \ref{lvdc}}, let $f(n)=\frac{(a_{1}-a_{2})n^{2}+2(\nu-a_{1}\tau)n}{2N}$ or $f(n)=\frac{(a_{1}-a_{2})n^{2}+2(\nu-a_{2}\tau)n}{2N}$, $\rho=\frac{a_{1}-a_{2}}{N}$, $\alpha=1$ and $\xi=N-|\tau|$. Substituting them into (\ref{lem5}), we obtain (\ref{T4}).
\end{IEEEproof}

For sufficiently large $N$ and $a_1-a_2\ll N$, the upper bound (\ref{T4}) of the maximum CAF magnitude for $\mathbf{s}^{1}$ and $ \mathbf{s}^{2}$ is in the order of $\mathcal{O}(\sqrt{N})$.

Based on \textit{Theorem \ref{T3ss}} and \textit{Theorem \ref{T4Css}}, the following corollary can be obtained.

\begin{corollary}
\label{ppchu}
     The Chu sequence set $(N,M,\theta_{\text{max}},\bm \Gamma)\text{-}\mathcal{A} = \left \{ \mathbf{s}^{m}\right \} _{m=1}^{M}$ is an order-optimal sequence set for aperiodic AF in LAZ $\bm \Gamma$ with $Z_X< \left \lfloor \frac{N}{\max_{m}{|a_m|}}-1 \right \rfloor $ and $Z_y\leq \min_{m}{|a_m|}$, if $N\geq5\max_{m}{|a_m|}$ and $\max_{m_1,m_2 \in [1,M]}\left \{{a_{m_1}}-{a_{m_2}}\right \}\ll N$.
\end{corollary}

\section{Discussions and Comparisons}
In this section, we evaluate the tightness of the proposed aperiodic AF lower bounds from two perspectives. We first demonstrate that it is tighter than the existing AF lower bound, and then we show the asymptotic achievability of our proposed AF lower bounds.

\subsection{Comparisons with the Existing AF Lower Bound}
The only existing aperiodic AF lower bound in the literature can be expressed as:
\begin{lemma} [{Existing Bound \cite[Theorem 4]{Ye2022}}]
    For any LAZ $\bm \Gamma$ ($1< Z_{x} \le N$ and $1\le Z_{y} \le N$),
    \begin{align}
    \label{p4f}
        {\theta}_{\text{max}}^{2} \geq N^2\frac{MZ_{x}Z_{y}-N-Z_x+1}{(N+Z_x-1)(MZ_{x}-1)Z_{y}}.
    \end{align}
\end{lemma}

\begin{IEEEproof}
    In \cite{Ye2022}, the proof for (\ref{p4f}) is an extension from the periodic case and is complicated. Here, we provide a concise proof using Welch's inner product method\cite{Welch1974}. For a  sequence set $\left \{\bm {\mu}^{m}\right \} _{m=1}^{\hat{M}}$ with $\hat{M}$ sequences of length $\hat{N}$ and energy $E$, their correlations satisfy \cite{Welch1974}
    \begin{align}
    \label{ipm}
    \sum_{m=1}^{\hat{M}}\sum_{m'=1}^{\hat{M}}\left |R_{\bm{\mu}^{m}, \bm{\mu}^{m'}}(0) \right |^{2}\geq \frac{(\hat{M}E)^{2}}{\hat{N}}.
    \end{align}
    Based on (\ref{ipm}) and \textit{Lemma~\ref{Lm1}}, let $\overline{\bm{x}}^{m, r} =\left [\widetilde{\bm{x}}^{m, r},\bm{0}_{1\times Z_x}\right ]$, since $\overline{\bm{x}}^{m, r}$ is of length $\hat{N}=N+Z_x-1$ and energy $E={N}$, we have
    \begin{align}
        \label{P3P}&MZ_{x}Z_{y}\left (MZ_{x}Z_{y}-Z_{y} \right){\theta}_{\text{max}}^{2} + N^2MZ_{x}Z_{y} \nonumber \\
         \geq& \sum_{m,m'=1}^{M}\sum_{i,i'=0}^{Z_{x}-1}\sum_{r,r'=0}^{Z_{y}-1} \left |R_{\mathbf c \left(\overline{\bm {x}}^{m, r}\right)_{i}, \mathbf c\left(\overline{\bm {x}}^{m', r'}\right)_{i'}} (0)\right |^{2} \nonumber \\
        \geq&  \frac{(MZ_{x}Z_{y}N)^{2}}{N+Z_x-1}.
    \end{align}
    The bound (\ref{p4f}) follows from (\ref{P3P}).
\end{IEEEproof}

\begin{table*}[htbp]
\label{table1}
\centering
\caption{Asymptotic AF Lower Bound Comparison (in the LAZ with $Z_x = N/4$ and $Z_y = 10$)}
\resizebox{1.95\columnwidth}{!}{
\begin{tabular}{|c|c|c|c|c|c|}
\hline
\textbf{AF Lower Bound} & \textbf{$M = 1$} & \textbf{$M = 2$} & \textbf{$M = 3$} & \textbf{$M = 4$} & \textbf{$M \geq 5$} \\ \hline
\cite[Theorem 4]{Ye2022}\textsuperscript{1} & $\theta^2_{\text{max}} \gtrsim 0.4000N$ & $\theta^2_{\text{max}} \gtrsim 0.6000N$& $\theta^2_{\text{max}} \gtrsim 0.6667N$ & $\theta^2_{\text{max}} \gtrsim 0.7000N$ & $\theta^2_{\text{max}} \gtrsim \frac{2N \left({10MN} - {5N} + 4\right)}{5 \left(5N - 4\right) \left(MN - 4\right)}N
$ \\ \hline
\textit{Corollary \ref{COROA}}\textsuperscript{2} & $\theta^2_{\text{max}} \gtrsim 0.6349N$ & $\theta^2_{\text{max}} \gtrsim 0.7418N$& $\theta^2_{\text{max}} \gtrsim 0.7892N$ & $\theta^2_{\text{max}} \gtrsim 0.8174N$ & $\theta^2_{\text{max}} \gtrsim \left(1-{\frac{2}{\sqrt{30M}}}\right)N$ \\ \hline
\textit{Corollary \ref{COROB}}\textsuperscript{3} & $\theta^2_{\text{max}} \gtrsim 0.6488N$ & $\theta^2_{\text{max}} \gtrsim 0.7516N$ & $\theta^2_{\text{max}} \gtrsim 0.7972N$ & $\theta^2_{\text{max}} \gtrsim 0.8244N$ & $\theta^2_{\text{max}} \gtrsim \left(1-{\frac{\pi}{\sqrt{80M}}} \right)N$ \\ \hline
\end{tabular}
}
\footnotesize
\begin{tabbing}
\textsuperscript{1} \= shown in (\ref{p4f});\\
\textsuperscript{2} \= shown in (\ref{C3}), from the weight vector $\mathbf w^A$, note that $Z_{x} > \sqrt{\frac{3N^{2}}{MZ_{y}}}$ and $MZ_{y}\geq 3$;\\
\textsuperscript{3} \= shown in (\ref{C5}), from the weight vector $\mathbf w^B$, note that $Z_x > \frac{\pi}{\gamma}$ and $5 \leq MZ_{y}\leq N^2$.\\
\end{tabbing}
\end{table*}

We first compare the results of our proposed bound with the benchmark bound (\ref{p4f}) when $N$ is sufficiently large:
\begin{observation}
    \label{ppcomp}
    For $M$, $Z_x$, $Z_y$ satisfying $Z_{x} > \sqrt{\frac{3N^{2}}{MZ_{y}}}$ and $MZ_{y}\geq 3$ or $Z_x > \frac{\pi}{\gamma}$ and $5 \leq MZ_{y}\leq N^2$, the respective asymptotic lower bound of our proposed method is tighter than that of \cite[Theorem 4]{Ye2022}.
\end{observation}

\textit{Observation \ref{ppcomp}} can be easily verified through numerical calculations, with partial results in Table I. One can observe that both our proposed AF bounds are tighter than the bound \cite[Theorem 4]{Ye2022} for different values of $M$. Additionally, note that the effective ranges of the bounds in \textit{Corollary \ref{COROA}} and \textit{Corollary \ref{COROB}} differ, with the bound in \textit{Corollary \ref{COROB}} being slightly tighter when both bounds are applicable.

We further compare our proposed bounds with the benchmark bound (\ref{p4f}) \cite{Ye2022} through two examples. 

\begin{figure*}[htbp]
\centering
\begin{tabular}{c}
\begin{subfigure}[b]{0.49\textwidth}
\centering
\includegraphics[trim=0cm 0.8cm 3.5cm 0.88cm, clip=true, width=\linewidth]{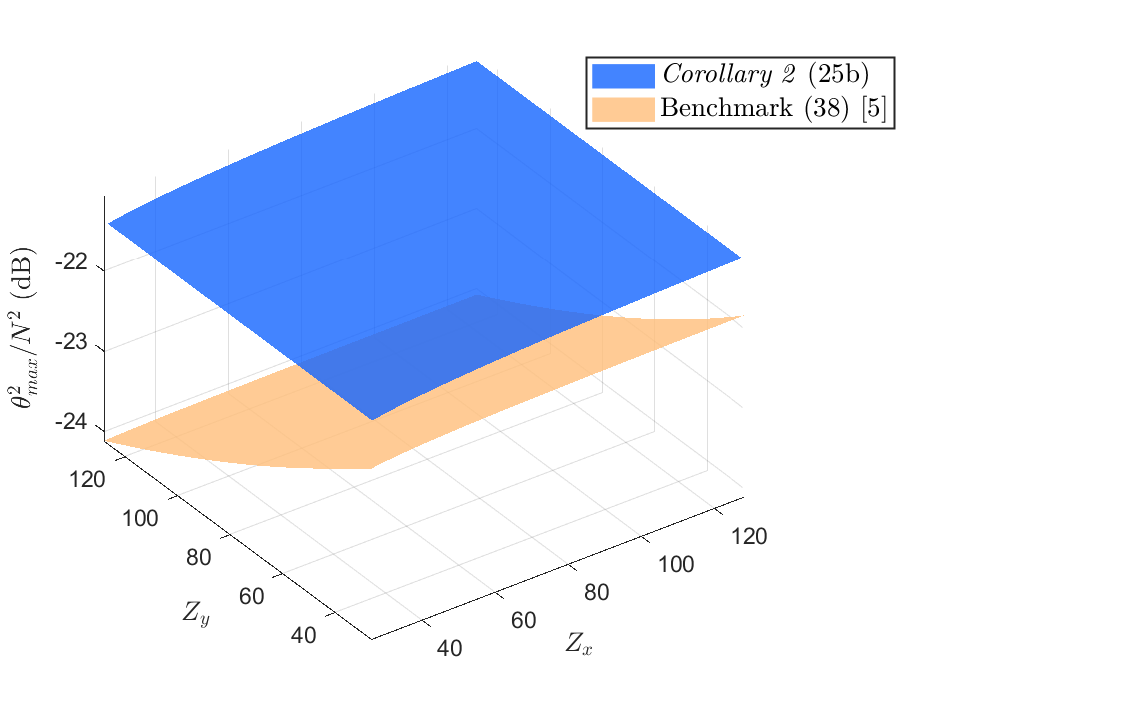}
\caption{Proposed bound versus the LAZ (with $N=128$ and $M=6$).\\ \;}
\label{f1a}
\end{subfigure}
\begin{subfigure}[b]{0.49\textwidth}
\centering
\includegraphics[trim=0.1cm 0.1cm 0.1cm 0.68cm, clip=true, width=\linewidth]{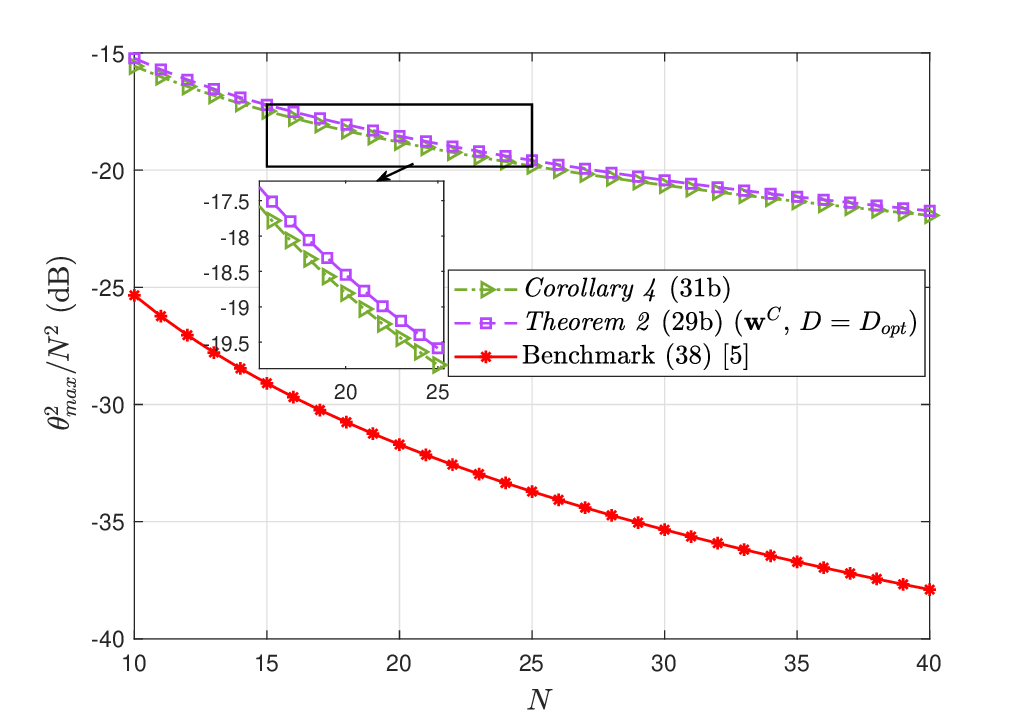}
\caption{Proposed bounds for LAZ versus sequence length $N$\\ (with $M=1$, $Z_x=N$ and $Z_y=2$).}
\label{f1b}
\end{subfigure}
\end{tabular}
\caption{Comparison of the proposed aperiodic AF lower bounds and the benchmark bound.}
\label{AFbound}
\end{figure*}

\textit{Example 1}: In Fig.\ref{f1a}, for LAZ with different $Z_x$ and $Z_y$, with $N=128$ and $M = 6$, one can observe that our proposed bounds (under the weight vector $\mathbf{w}^{B,q}$ with $q = \min\left \{{Z_x,\left \lfloor \frac{\pi}{\gamma} \right \rfloor+1}\right \}$) are tighter than the benchmark bound (\ref{p4f}) \cite{Ye2022}. As analyzed in Subsection {\ref{DDLAZB}}, the configuration of this example satisfies $q < N-\left \lfloor N\sqrt{\text{AFB}_{\text{ref}}} \right \rfloor$, hence the consideration of $D_{\text{opt}}$ is unnecessary. Furthermore, the proposed bounds exhibit a corresponding increase as $Z_x$ or $Z_y$ increases, which is a distinct behavior compared to the benchmark (\ref{p4f}). This also indicates that our proposed bounds align more with the physical significance. The rationale behind this is that achieving lower (non-trivial) AF magnitude is increasingly challenging for a larger LAZ. Therefore, it is logical that the AF lower bound should increase with the enlargement of the LAZ.

\textit{Example 2}: In Fig.\ref{f1b}, an example of the LAZ with $Z_x=N$ is presented with $M = 1$ and $Z_y=2$. This example meets the conditions outlined in \textit{Remark~\ref{R2}}, implying that $\mathbf{w}^{C}$ offers the tightest bound among all different weight vectors. Again, our proposed bounds are tighter than the benchmark bounds (\ref{p4f}). Furthermore, setting $D = D_{\text{opt}}\approx \left \lfloor N\sqrt{\widetilde{\text{AFB}}_{\text{ref}}} \right \rfloor$ effectively tightens this bound further, which shows the significance of the $D$ last delays consideration.

These examples again demonstrate that our proposed bounds are tighter than the only existing bound \cite[Theorem 4]{Ye2022}.

\subsection{Asymptotic Achievability of the Proposed Bounds}
In Section \ref{Chusection}, we have shown that our proposed AF lower bounds are asymptotically achievable in certain LAZ by the Chu sequences. In this subsection, we provide numerical examples to illustrate the closeness between the proposed aperiodic AF lower bounds and the maximum AAF and CAF magnitudes of Chu sequences, thereby demonstrating the tightness of our proposed bounds. 

\begin{figure*}[htbp]
\centering
\includegraphics[trim=0cm 0.11cm 0cm 0.18cm, clip=true, height =6.8cm]{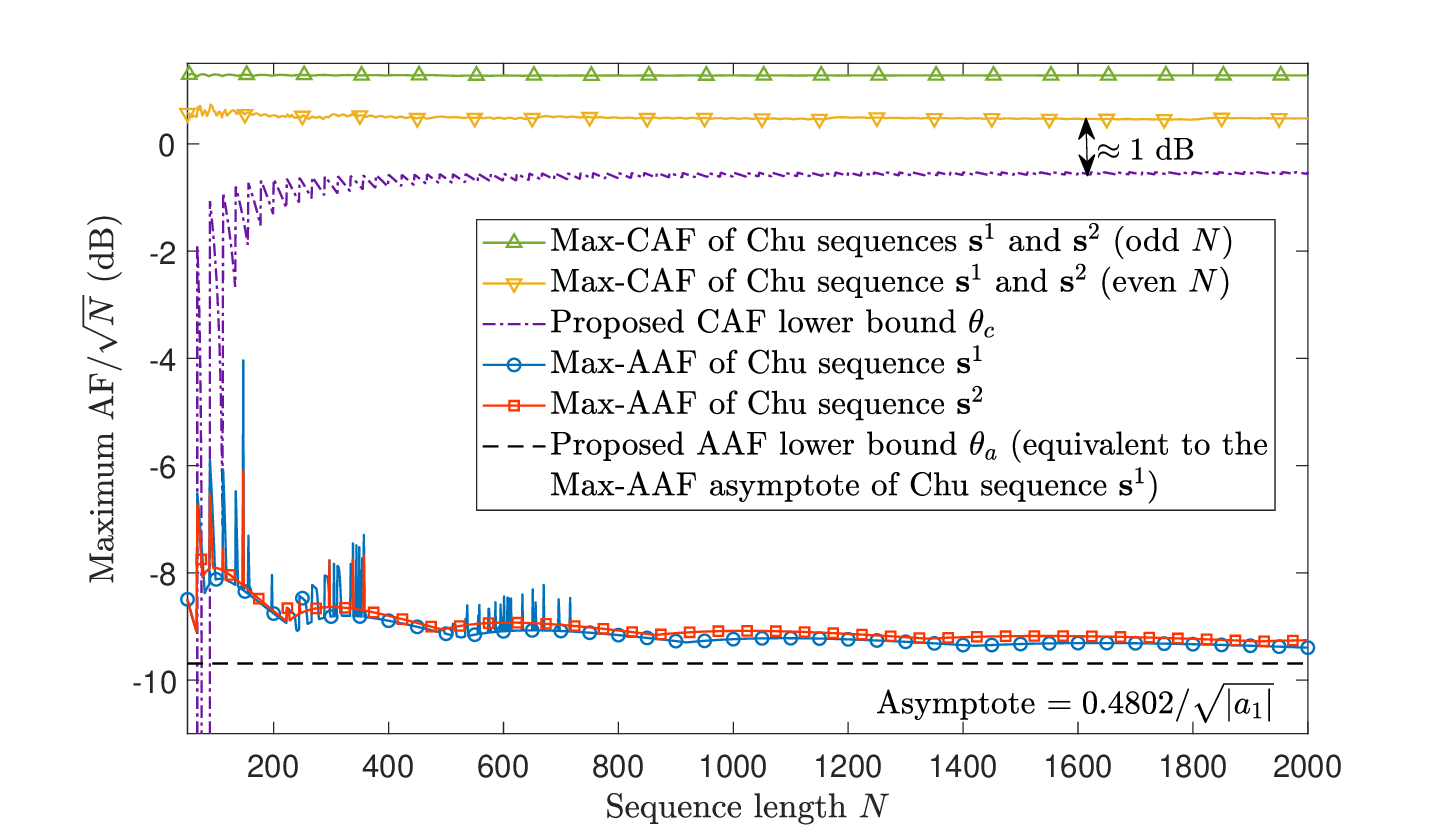}
\caption{Comparison of the maximum AAF/CAF magnitudes of a Chu sequence pair $\mathbf{s}^1$ and $\mathbf{s}^2$ ($a_1=20$, $a_2 = 19$) and the proposed aperiodic AAF/CAF lower bounds (\ref{C3}a) associated to the LAZ with $Z_x= \left \lfloor \frac{9N}{10|a_1|} \right \rfloor$ and $Z_y = a_2$.}
\label{f3}
\end{figure*}

In Fig.~\ref{f3}, we show the maximum AAF/CAF magnitudes of a Chu sequence pair $\mathbf{s}^1$ and $\mathbf{s}^2$ with $a_1=20$ and $a_2=19$ for different values of $N$, along with the proposed aperiodic AF lower bounds associated to the LAZ ($Z_x= \left \lfloor \frac{9N}{10|a_1|} \right \rfloor$ and $Z_y = a_2$ according to \textit{Corollary \ref{ppchu}}). We set $\theta_a = 0.4802\sqrt{\frac{N}{a_1}}$ as the asymptotic value of the maximum AAF magnitude for $\mathbf{s}^1$, which is smaller than that of $\mathbf{s}^2$. According to the tradeoff between $\theta_a$ and $\theta_c$ as shown in (\ref{C3}a), $\theta_c$ is represented by the purple line in the figure. Note that some values in Fig.~\ref{f3} exceed $0$~dB because $\sqrt{N}$ is used as the normalization factor here for calculating decibels instead of $N$.

One can observe that the maximum AAF magnitudes of both $\mathbf{s}^1$ and $\mathbf{s}^2$ asymptotically approach their respective asymptotic values, while the maximum CAF magnitude between them only slightly exceeds the bound $\theta_c$ when $N$ is sufficiently large. Despite a certain gap, this demonstrates that the proposed bounds are quite tight for the aperiodic case, and their order is asymptotically achievable. Similarly, other Chu sequence pairs satisfying the conditions in \emph{Corollary \ref{ppchu}} also exhibit maximum AF magnitudes that closely approach the proposed bounds, and such order-optimal sequence pairs are readily extendable to form order-optimal sequence sets.

\section{Conclusions}
In this work, we have derived generalized Arlery-Tan-Rabaste-Levenshtein AF lower bounds with respect to the LAZ in the delay-Doppler plane. We have also presented the trade-offs between the lower bounds for the aperiodic AAF and CAF. Our first innovation is the introduction of two weight vectors in order to characterize the individual influences of the delay and Doppler shifts, respectively. Moreover, our bounds are derived by exploring certain insights of the zero delay and $D$ last delays that are associated with the structural properties of aperiodic AF. Furthermore, it has been shown that our proposed AF lower bounds are asymptotically achievable by the Chu sequence sets and are tighter than the existing benchmark bound.




\appendices
\section{Proof of Corollary~\ref{COROA}}
\label{appA}

By substituting $\mathbf{w}^{A,q}$ and $D = 0$ into (\ref{T2}a), we have
     \begin{align}
         &{\theta}_{c}^{2} (M-1)+{\theta}_{a}^{2} \left (1-\frac{1}{q} \right) \nonumber
\\\geq& M\left(N-\sum_{s, t=0}^{Z_{x}-1}l_{s, t, N}w_{s}w_{t}-\frac{N^{2}}{MZ_{y}}\sum_{i=0}^{Z_{x}-1}  w_{i}^{2} \right ) \nonumber\\
= &M\left (N-\sum_{s, t=0}^{q-1}|s-t|\frac{1}{q^{2}}-\frac{N^{2}}{qMZ_{y}} \right ) \nonumber\\
= &M \left (N-\frac{(q-1)(q+1)}{3q}-\frac{N^{2}}{qMZ_{y}} \right).
     \end{align}
     Based on ${\theta}_{\text{max}}^{2} \left( M-\frac{1}{q}\right) \geq {\theta}_{c}^{2} (M-1)+{\theta}_{a}^{2} \left (1-\frac{1}{q} \right)$, we have ${\theta}_{\text{max}}^{2}  \geq \frac{3qMNZ_{y}-3N^{2}-q^{2}MZ_{y}+MZ_{y} }{3(qM-1)Z_{y}}$.
     
    When $MZ_{y}\geq 3$, there is an integer $\hat{q}$ satisfies $1\leq \hat{q} \leq N$, and $\hat{q}= \sqrt{\frac{3N^{2}}{MZ_{y}}}-\epsilon$, where $0 \leq \epsilon < 1$,
    \begin{align}
        &{\theta}_{c}^{2} (M-1)+{\theta}_{a}^{2} \left (1-\frac{1}{\sqrt{\frac{3N^{2}}{MZ_{y}}}} \right) \nonumber\\ \geq&{\theta}_{c}^{2} (M-1)+{\theta}_{a}^{2} \left (1-\frac{1}{\hat{q}} \right)
\nonumber\\\geq& M\left(N-\frac{\hat{q}^{2}-1}{3q}-\frac{N^{2}}{\hat{q}MZ_{y}}\right) \nonumber \\\geq& M\left (N-\frac{2N}{\sqrt{3MZ_{y}}} \right ).
    \end{align}

\section{Proof of Theorem~\ref{T3ss}}
\label{appB}

Since $A_{\mathbf{s}}(\tau, \nu) = {A^{\ast}_{\mathbf{s}}(-\tau, -\nu)}$, we only need to consider $\tau >0$ to analyze the maximum AF magnitude in a given LAZ. Note that if $\nu-a\tau \neq uN$ for any integer $u$, $\sum_{n=0}^{N-1} e^{\frac{j2\pi(\nu-a\tau)n}{N}}=0$ \cite{Sarwate1979}. For $\tau>0$, we have
\begin{align}
\label{Chuaafrl}
\left |A_{\mathbf{s}}(\tau, \nu) \right | & =\left |\sum_{n=0}^{N-1-\tau} {s}_{n}^{a}{\left({s}_{n+\tau}^{a}\right)}^{\ast}e^{\frac{j2\pi\nu n}{N}} \right | \nonumber\\
&=\left |\sum_{n=0}^{N-1-\tau} e^{\frac{j\pi an^{2}}{N}}e^{\frac{-j\pi a(n+\tau)^{2}}{N}}e^{\frac{j2\pi \nu n}{N}} \right | \nonumber\\
& =\left|\sum_{n=0}^{N-1-\tau} e^{\frac{j2\pi (\nu - a\tau) n}{N}}\right| 
\nonumber\\
&=\left|\sum_{l=0}^{N-1-(N-\tau)} e^{\frac{-j2\pi (\nu + a(N-\tau)) l}{N}}\right|\nonumber\\
& =\left|A_{\mathbf{s}}(N-\tau, -\nu)\right|.
\end{align}
Based on (\ref{Chuaafrl}), one can easily verify that $A_{\mathbf{s}}(\tau, \nu) = A_{\mathbf{s'}}(\tau, -\nu)$ where $\mathbf{s}'=\left [{s}^{-a}_{0}, {s}^{-a}_{1}, \cdots, {s}^{-a}_{N-1}\right ]$. Thus, in the following, we only need to consider that $a<-1$, $a< \nu <-a$ and $0<\tau \leq\frac{\beta N}{|a|}-1$, $\frac{1}{2}<\beta<1$, so we have $0<\nu-a\tau <\beta N$. We consider the following two cases:

Case 1: $1 \leq \tau\leq 0.45\sqrt{\frac{N}{|a|}}.$

From the choice of $a$, $\nu$ and $\tau$, it is clear that $\nu-a\tau \neq uN$, holds for any integer $u$. By (\ref{Chuaafrl}), \textit{Lemma \ref{Lm2}} and for $1 \leq \tau\leq 0.45\sqrt{\frac{N}{|a|}}$, we have 
\begin{align*}
    \left |A_{\mathbf{s}}(\tau, \nu) \right |& = \left |A_{\mathbf{s}}(N-\tau, -\nu) \right | \leq N-(N-\tau)=\tau\nonumber\\
&\leq 0.45\sqrt{\frac{N}{|a|}}.
\end{align*}
Thus
\begin{align}
\label{C1R}
    \frac{\left |A_{\mathbf{s}}(\tau, \nu) \right |}{\sqrt{N}} \leq \frac{ 0.45}{\sqrt{|a|}},\quad \text{for Case 1}.
\end{align}

Case 2: $0.45\sqrt{\frac{N}{|a|}}< \tau\leq \frac{\beta N}{|a|}-1$. 

Since $N\geq 5|a|$, we have 
\begin{align*}
    &0<-|a|+1+0.45\sqrt{5}|a|\\&\leq-|a|+1+0.45\sqrt{|a|N}<\nu-a\tau <\beta N.
\end{align*}
Note that
\begin{align}
\label{ChuAAFs}
&\left |A_{\mathbf{s}}(\tau, \nu) \right | \nonumber\\ =&\left |\sum_{n=0}^{N-1-\tau} e^{\frac{j2\pi (\nu-a\tau)n}{N}} \right | =\left |\frac{1-e^{\frac{j2\pi (\nu-a\tau)(N-\tau)}{N}}}{1-e^{\frac{j2\pi (\nu-a\tau)}{N}}} \right | \nonumber\\
 =&\sqrt{\frac{1-\cos \frac{2\pi(\nu-a\tau)(N-\tau)}{N}}{1-\cos \frac{2\pi(\nu-a\tau)}{N}}} =\sqrt{\frac{1-\cos \frac{2\pi(\nu-a\tau)\tau}{N}}{1-\cos \frac{2\pi(\nu-a\tau)}{N}}} \nonumber \\
 =&\left |\frac{\sin\frac{\pi(\nu-a\tau)\tau}{N}}{\sin \frac{\pi(\nu-a\tau)}{N}}\right |.
\end{align}
Then, we divide Case 2 into the following two sub-cases:

Sub-case 2-1: $0.45\sqrt{\frac{N}{|a|}}< \tau\leq \frac{\beta N}{|a|}-1$ with $\nu-a\tau\geq \frac{\beta}{\sin (\beta \pi)}\sqrt{5|a|N}$. 

Since $\beta<1$, $\sin x$ is concave downward in $[0, \beta \pi]$. Thus $\sin x > \frac{\sin (\beta \pi)}{\beta \pi} x>0$, for $0<x< \beta \pi$. Since $0<\nu-a\tau<\beta N$, we have $0<\frac{\pi(\nu-a\tau)}{N}<\beta \pi$. Hence 
\begin{align*}
    \sin\frac{\pi(\nu\!-\!a\tau)}{N}\!>\!\frac{\sin (\beta \pi)}{\beta \pi}\!\cdot\!\frac{\pi(\nu\!-\!a\tau)}{N}\!=\!\frac{\sin (\beta \pi)}{\beta N}\!\cdot\! (\nu\!-\!a\tau)\!>\!0.
\end{align*}
Then based on (\ref{ChuAAFs}), we have 
\begin{align}
\label{C21R}
    &\frac{\left |A_{\mathbf{s}}(\tau, \nu) \right |}{\sqrt{N}}\nonumber\\=&\frac{\left |\sin\frac{\pi(\nu-a\tau)\tau}{N}\right |}{\sqrt{N}\left |\sin \frac{\pi(\nu-a\tau)}{N}\right |}<\frac{1}{\sqrt{N}\frac{\sin(\beta \pi)}{\beta N}\cdot(\nu-a\tau)}\nonumber\\=&\frac{\beta}{\sin (\beta \pi)}\cdot\frac{\sqrt{N}}{(\nu-a\tau)} \nonumber\\
    \leq& \frac{\beta}{\sin (\beta \pi)}\cdot\frac{\sin (\beta \pi)}{\sqrt{5|a|}\beta} = \frac{1}{\sqrt{5|a|}}, \quad \text{for Sub-case 2-1}.
\end{align}

Sub-case 2-2: $0.45\sqrt{\frac{N}{|a|}}< \tau\leq \frac{\beta N}{|a|}-1$ with $\nu-a\tau< \frac{\beta}{\sin (\beta \pi)}\sqrt{5|a|N}$.

In this sub-case, $\frac{\nu-a\tau}{N} \rightarrow 0$ as $N \rightarrow +\infty$. Following (\ref{ChuAAFs}), for $N \rightarrow +\infty$, we have
\begin{align}
\label{SC22}
    {\left |A_{\mathbf{s}}(\tau, \nu) \right |}^2&\rightarrow \frac{1-\cos \frac{2\pi(\nu-a\tau)\tau}{N}}{\frac{1}{2} \left (\frac{2\pi(\nu-a\tau)}{N} \right )^{2}}=\frac{1-\cos \frac{2\pi(\nu-a\tau)\tau}{N}}{\frac{\pi(\nu-a\tau)}{N\tau}\cdot \frac{2\pi(\nu-a\tau)\tau}{N}}\nonumber \\
    &=\frac{N\tau}{\pi(\nu-a\tau)}F\left(\frac{2\pi(\nu-a\tau)\tau}{N}\right ),
\end{align}
where $F(\varphi)=\frac{1-\cos\varphi}{\varphi}$ with
\begin{align*}
    \varphi=&\frac{2\pi(\nu-a\tau)\tau}{N}=\frac{\tau}{\nu-a\tau}\frac{2\pi(\nu-a\tau)^{2}}{N}\\<&\frac{2\pi\tau}{\nu-a\tau}\cdot\frac{5\beta^{2}|a|}{\sin^{2} (\beta \pi)}.
\end{align*}
For $\varphi>2\pi$, $F(\varphi)\leq \frac{2}{\varphi} \leq \frac{1}{\pi}\approx 0.3183$. For $0<\varphi\leq 2\pi$, the maximum value of $F(\varphi)$ equals $F(\varphi_{0})\approx 0.7246$, $\varphi_{0}\approx 2.3311$ \cite{Mow1997}.

Then, we consider the $\frac{\tau}{\nu-a\tau}$ part in (\ref{SC22}). Note that $\frac{\tau}{\nu-a\tau}=\frac{1}{\frac{\nu}{\tau}-a}$, $0.45\sqrt{\frac{N}{|a|}}< \tau\leq \frac{\beta N}{|a|}-1$, and $a< \nu <-a$, we have
\begin{align*}
    \frac{1}{\frac{-a-1}{0.45\sqrt{N/|a|}}-a} \leq\frac{\tau}{\nu-a\tau}\leq \frac{1}{\frac{a+1}{\beta N/|a|-1}-a}. 
\end{align*}
For $N\rightarrow+\infty$, 
\begin{align*}
    \frac{1}{\frac{-a-1}{0.45\sqrt{N/|a|}}-a} \rightarrow \frac{1}{|a|}\quad \text{and} \quad\frac{1}{\frac{a+1}{\beta N/|a|-1}-a}\rightarrow \frac{1}{|a|},
\end{align*}
so $\frac{\tau}{\nu-a\tau}\rightarrow \frac{1}{|a|}$. Thus, from (\ref{SC22}), we have
\begin{align}
\label{C22R}
    &\lim_{N \to +\infty}  \frac{\max_{(\tau,\nu)} {\left |A_{\mathbf{s}}(\tau, \nu) \right |}}{\sqrt{N}} = \sqrt{\frac{F(\varphi_{0})}{\pi |a|}}  =\frac{0.4802}{\sqrt{|a|}},\nonumber\\& \text{for Sub-case 2-2}.
\end{align}

Overall, based on the results from (\ref{C1R}), (\ref{C21R}), and (\ref{C22R}), we conclude (\ref{TC3}).



\bibliographystyle{IEEEtran}
\bibliography{BibOne.bib}

\end{document}